# UVMag: Space UV and visible spectropolarimetry

Martin Pertenais*[a,b,c], Coralie Neiner[c], Laurent Parès[a,b], Pascal Petit[a,b], Frans Snik[d], Gerard van Harten[d]

[a]Université de Toulouse ; UPS-OMP ; IRAP Toulouse, France
[b]CNRS ; IRAP ; 14 av Edouard Belin F-31400 Toulouse, France
[c]LESIA, Observatoire de Paris, CNRS UMR 8109, UPMC, Université Paris Diderot, 5 place Jules Janssen, 92190 Meudon, France
[d]Leiden Observatory, Leiden University, P.O. Box 9513, 2300 RA Leiden, The Netherlands
*martin.pertenais@irap.omp.eu

## ABSTRACT

UVMag is a project of a space mission equipped with a high-resolution spectropolarimeter working in the UV and visible range. This M-size mission will be proposed to ESA at its M4 call. The main goal of UVMag is to measure the magnetic fields, winds and environment of all types of stars to reach a better understanding of stellar formation and evolution and of the impact of stellar environment on the surrounding planets. The groundbreaking combination of UV and visible spectropolarimetric observations will allow the scientists to study the stellar surface and its environment simultaneously.

The instrumental challenge for this mission is to design a high-resolution space spectropolarimeter measuring the full-Stokes vector of the observed star in a huge spectral domain from 117 nm to 870 nm. This spectral range is the main difficulty because of the dispersion of the optical elements and of birefringence issues in the FUV. As the instrument will be launched into space, the polarimetric module has to be robust and therefore use if possible only static elements. This article presents the different design possibilities for the polarimeter at this point of the project.

## 1. SCIENCE CASE FOR UVMAG

Important insights into the formation and evolution of all types of stars can be obtained through the measurement of their spectra in the UV and visible domains. The visible spectrum allows us to gain information about the surface of the star itself, while the UV domain is crucial because it is very rich in atomic and molecular lines, contains most of the flux of hot stars and the signatures of the stellar environment (e.g. of the chromosphere). In addition, performing UV and visible spectroscopy over a full stellar rotation period will allow us for the first time to reconstruct the 3D maps of stars and their environment simultaneously. Adding polarimetric power to the spectrograph will multiply tenfold the capabilities of extracting information on stellar magnetospheres, winds, disks, and magnetic fields.

In particular, with UVMag it will be possible to follow the life cycle of matter in the Galaxy. For example we will investigate the role of magnetic fields in accretion and mass loss along stellar evolution. This can be done by probing fossil magnetic fields in hot stars and dynamo fields in cool stars, from the pre-main sequence (T Tauri stars, Herbig Ae/Be stars…) to late stages of evolution (giants, supergiants, Wolf-Rayet stars, Luminous Blue Variables…). Moreover, we will investigate the structure, geometry and dynamics of stellar environments across stellar evolution, e.g. disks around hot stars such as the decretion disks of classical Be stars, chromospheres and the lower corona of cool stars, the wind of hot and cool stars, and colliding winds between two components of multiple systems. In the case of binary stars, it will also be possible to investigate the mass and angular momentum transfers, which strongly modify their evolution. In addition, dying stars directly impact their environment by providing radiation and chemical elements to the interstellar medium (ISM) from which next generations of stars and planets then form. In particular, massive stars end their life as supernovae, which serve as the mechanism for the creation of heavy elements and for their dispersal in the ISM. Life on the Earth would not be possible without the occurrence of supernovae before the formation of our Sun. This also has strong consequences on the chemical evolution of our Galaxy and on the evolution of its energy distribution.

Moreover, the changing stellar UV radiation and magnetic fields affect the formation of planets around stars and the emergence of life on these exoplanets. Thanks to UVMag we will study the interaction between stars and their planets, in particular magnetospheric interactions and tides. This will allow us to study the environmental conditions for the emergence of life on rocky planets. Indeed, we know that the Earth's magnetosphere protects us from the flares and coronal mass ejections (CME) from the Sun and that UV irradiation can be fatal for life. By studying other stellar systems, we will be able to better understand our own.

To reach our science objectives, we are designing UVMag, a M-size space mission equipped with a high-resolution spectropolarimeter working in the UV and visible spectral range. UVMag will cast new light onto stellar physics by addressing the exciting and important topics listed above and many more.

## 2. REQUIREMENTS

These various science cases lead to some requirements on the performances of the instrument. Particularly it will influence the development of the polarimetric and spectroscopic instruments.

Table 1. Requirements details for the UVMag instrument

| Specification | Minimum requirement | Goal |
|---|---|---|
| Spectral range | [117 nm; 320 nm] + [390 nm; 870 nm] | [90 nm; 1000 nm] |
| SNR | 100 | 200 |
| Target magnitude | V=3-10 | V=2-15 |
| Exposure time (mean) | 20 min | |
| UV Resolution | 25000 | 100000 |
| Visible Resolution | 35000 | 80000 |
| $V_{rad}$ accuracy | 1 km.s$^{-1}$ | 0.3 km.s$^{-1}$ |
| Instru. Polarization | <3% | <1% |

The spectral resolution is defined by the ratio of the wavelength over the resolved spectral element (over 2 pixels), i.e.

$$R = \frac{\lambda}{\Delta\lambda} \tag{1}$$

The first information that can be extracted from these requirements is the size of the telescope needed. Using the SNR wanted in every spectral element given by the resolution, the diameter of the telescope needed to observe a 10-magnitude star during 20 min is determined. This leads to a telescope diameter of around 1.3 m.

The accuracy of the radial velocity is directly linked to the thermo-mechanical stability of the spectrometer. It determines the precision obtained in the positioning of several spectral lines and the stability of these positions. This requirement could have been given in wavelength units, but converting it into a velocity allows us to have a value independent from the observation wavelength.

Finally, minimizing the instrumental polarization is a key in the design of the whole instrument. This will affect the capacity to measure weak polarization information. In fact, each optical element in the instrument will modify the entrance polarization of the light that we want to measure. The modification of the polarization state by some optics in the instrument is called the instrumental polarization. The challenge is to minimize it for every element on the optical path before the polarimeter. Particularly, the most important is to minimize the polarization cross-talk effect between Q, U and V, i.e. for example a linear entrance polarization changed into circular polarization.

## 3. POLARIMETER DESIGN

### 3.1 Introduction

The polarimeter is really the key component of the instrument and a polarimeter working from the FUV to the NIR has never been built in the past. This huge spectral range prevents us to use a regular polarimeter as the ones from some existing ground based spectropolarimeter, such as ESPaDOnS at CFHT [1] or Narval at TBL at the Pic du Midi [2].

A polarimetric device is made of two core elements: a modulator and an analyzer. Generally the modulator is made of one or several rotating birefringent waveplates or Fresnel rhombs, i.e. the polarization state is temporally modulated by the rotation of the plates. A birefringent plate will add a phase retardance between the two orthogonal polarization projections of the entrance light, for example a quarter wave plate will make a phase difference of a quarter of a wavelength, i.e. add a phase delay of $\pi/2$ between the two polarization states. However this delay is strongly dependent on the wavelength because the crystals used are strongly dispersive and, as a consequence, their birefringence is fully chromatic. For this reason, this classical method is not applicable in our case. The modulator has to be followed by a polarization analyzer, also called polarizer. This element will separate spatially the two polarization states to allow the measurement afterwards by a detector. The use of a Wollaston-type polarizer is useful in order not to waste flux, because it makes the simultaneous measurement of the two states possible, and not just one of them.

### 3.2 Material

The modulator has to be made with a birefringent material. The difficulty for this project is to find a material that is transparent and birefringent for the whole spectral range, especially in the FUV. The only material available fulfilling these criteria is the Magnesium Fluoride $MgF_2$. The cut-off wavelength for the transparence of $MgF_2$ is around 115 nm and its birefringence $\Delta n = n_0 - n_e$ can be computed using the measurements of [3], [4] and [5].

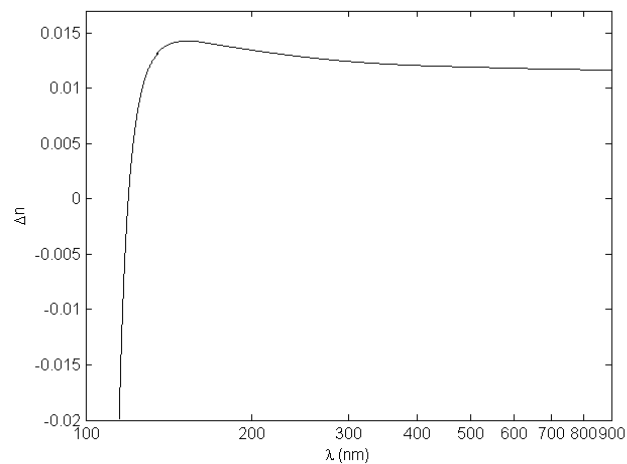

Figure 1. Interpolated curve of the birefringence of the Magnesium Fluoride over the wavelength range of interest: [115 nm; 870 nm] extracted from [3], [4] and [5].

We can see in Figure 1 that the birefringence of the MgF$_2$ is almost constant over a large spectrum from 145 nm to 900 nm. However the value of Δn is dramatically decreasing in the FUV, drops to 0 at 119.49 nm and then reverses its sign. This drop-off of the birefringence will have to be taken into account in the simulations and later on in the calibration phase. The equations (2), (3) and (4) give the interpolated expression for the birefringence over a given spectral range with the wavelengths given in nanometers.

From 115 nm to 135 nm:

$$\Delta n = 2.6986132984 \cdot 10^{-8} \cdot \lambda^5 - 1.7347731841 \cdot 10^{-5} \cdot \lambda^4 + 4.4620467721 \cdot 10^{-3} \cdot \lambda^3 \\ - 0.57406231101 \cdot \lambda^2 + 36.945138989 \cdot \lambda - 951.61055835 \quad (2)$$

From 135 nm to 160 nm:

$$\Delta n = 2.2842837414 \cdot 10^{-10} \cdot \lambda^5 - 1.7411445238 \cdot 10^{-7} \cdot \lambda^4 + 5.3125759992 \cdot 10^{-5} \cdot \lambda^3 \\ - 8.1126282732 \cdot 10^{-3} \cdot \lambda^2 + 0.62014682992 \cdot \lambda - 18.973915883 \quad (3)$$

From 160 nm to 900 nm:

$$\Delta n = -8.910586265 \cdot 10^{-17} \cdot \lambda^5 + 2.8705713841 \cdot 10^{-13} \cdot \lambda^4 - 3.6375402188 \cdot 10^{-10} \cdot \lambda^3 \\ + 2.2796309452 \cdot 10^{-7} \cdot \lambda^2 - 7.1945365121 \cdot 10^{-5} \cdot \lambda + 0.021205896742 \quad (4)$$

Another way to interpolate the birefringence curve is to determine dispersion equations respectively for the ordinary and the extraordinary refractive indexes of the material. To do this, the Sellmeier equation (5) is used where λ is the wavelength in microns and $A_i$, $L_i$ coefficients dependent on the material and shown in the Table 2 for the MgF$_2$. The birefringence curve will then be the difference, but the result is not as close to the actual curve as with the previous equations.

$$n^2 - 1 = \sum_i \frac{A_i \lambda^2}{(\lambda^2 - L_i)} \quad (5)$$

Table 2. Sellmeier coefficients for the MgF$_2$ between 120nm and 900nm

|  | O ray | E ray |
|---|---|---|
| $A_1$ | 0.445890386 | 0.473834349 |
| $A_2$ | 0.443593421 | 0.448210181 |
| $A_3$ | 0.0602750747 | 142330.615 |
| $L_1$ | $1.35344324 \cdot 10^{-11}$ | $7.98906037 \cdot 10^{-4}$ |
| $L_2$ | $9.27979982 \cdot 10^{-3}$ | $8.9949434 \cdot 10^{-3}$ |
| $L_3$ | 8.49522427 | 15817393.2 |

The only alternative to avoid this birefringence curve would be to use a reflective polarimeter. Some techniques are already developed in this direction, see [6] and [7], but we present in this paper only systems by transmission, using MgF$_2$.

### 3.3 Combination of Spatial and Spectral Modulation

The first possibility to build the polarimetric modulator is to modulate the polarization spatially and spectrally at the same time with two orthogonal directions. The main advantage of such a system is that we will not have any moving element in the polarimeter. The idea is well explained in [8] and consists in creating a slit where the retardance is varying along the spatial direction x. The polarizer afterwards is a classical one and by dispersing this slit with the spectrometer we will obtain a spectrum with polarimetric modulation in the orthogonal direction of the dispersion. At this point we do not need to rotate a waveplate in order to obtain the polarization modulation because there is a continuum of polarization states over the slit. Figure 2 shows the principle in more details. To create this slit, we use two wedges made of MgF$_2$, where the second wedge is twice as thick as the first one and has a fast axis oriented 45° to the first one.

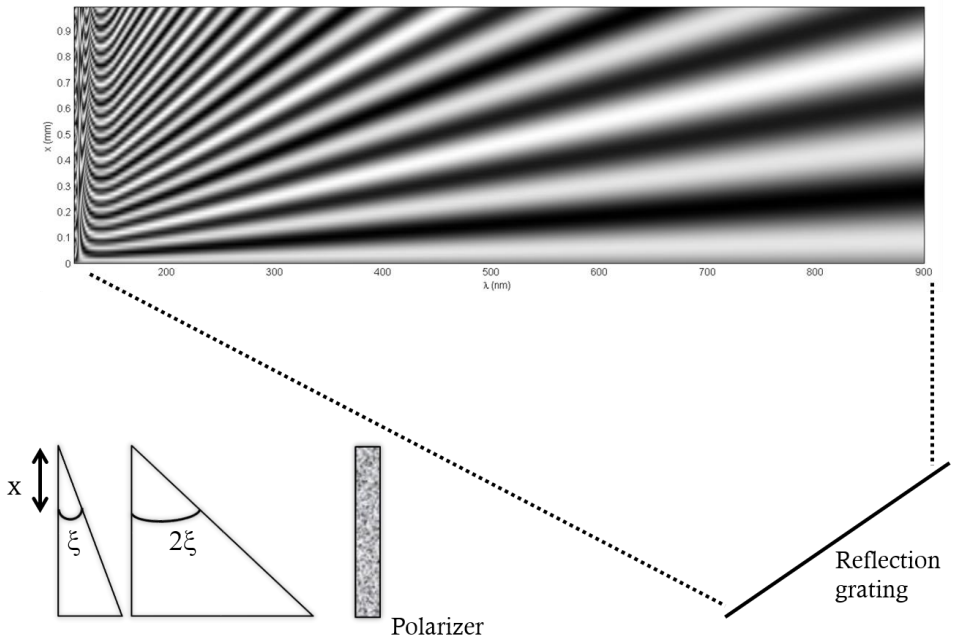

Figure 2. The two entrance wedges create the modulation of the polarization states, which are then selected by the polarizer. The grating disperses this slit to create the spectrum and the modulation of the polarization in the orthogonal direction of the dispersion.

As the retardance of a plate is directly linked to its thickness we can easily express the spatial height x as a function of the retardance, as shown in the equation (6) with Φ the retardance, ξ the wedge submit angle and Δn the birefringence.

$$x = \frac{\Phi}{6\pi \cdot \tan(\xi)} \cdot \frac{\lambda}{\Delta n(\lambda)} \qquad (6)$$

Considering the Mueller matrix of the 2 wedges and the one of the polarizer, the output intensity is determined as a function of the input Stokes vector we want to measure, of the retardance Φ given with the equation (6), and of the polarizer angle θ.

$$I_{out}(x,\lambda) = \frac{1}{2}(I + Q(\cos\Phi\cos 2\theta + \sin\Phi\sin 2\theta) + U\cos 2\Phi\sin 2\theta \\ + V(\cos\Phi\sin 2\Phi\sin 2\theta - \sin\Phi\cos 2\theta)) \qquad (7)$$

Equation (7) is plotted on the upper side of Figure 2 for every wavelength of interest. In fact, the measured intensity on the detector $I_{out}$ depends on $\Phi$ which depends on $\lambda$ and x. Performing a Fourier analysis of this equation shows that we have several modulation frequencies in this signal with periods of $2\pi$, $\pi$ and $2\pi/3$. To measure correctly the signal we have to sample it with a phase sampling of at least $\pi/3$ to satisfy the Nyquist criterion on the highest frequency. On the other hand, the lowest frequency has a period of $2\pi$, and it would be optimal to have at least 2 periods of this signal for a precise measurement. As a conclusion, we have to sample a signal modulation of $4\pi$, with a precision of $\pi/3$.

Because of the proportionality of the spatial direction x with the wavelength $\lambda$, the space gap to see the phase modulation is very low in the UV compared to in the visible, especially on the red side. This means that the space gap corresponding to a phase of $\pi/3$ in the UV range will define the sampling needed. As we use pixels of around 15μm side, the use of $\xi=2°$ gives a gap of 16 μm for $\pi/3$ phase retardance at 145 nm.

Using a wedge angle of 2°, requires a slit height of at least 1mm to be able to measure between 1 and 2 periods of the lowest frequency signal at 870 nm, where the distance x is the highest for a given phase delay $\Phi$.

These considerations define a first sizing of the instrument. The wedges will be around 1mm high with a submit angle of respectively 2° and 4°. A correct sampling of the slit in the UV requires a resolution of 15 μm, which is also the size of the pixels; this necessitates a magnification between the polarimeter and the detector (basically the one of the spectrometer) of 1.

This polarimeter concept is really interesting for UVMag, as it does not include any moving part and makes a complete measurement of the Stokes vector for each wavelength with a single shot. However, a first problem will be the size of the slit. The calculations give for the system a 1 mm slit with a magnification of the spectrometer of 1; this means that each order will occupy 1 mm on the detector. Considering the cross-disperser of the echelle-spectrometer that follows the polarimeter, this will lead to very large detectors. Furthermore, this concept has only been tested by Sparks et al. [8] and more elaborated laboratory tests have to be carried out to prove the feasibility and the precision in the polarization measured with such a system.

## 3.4 Polychromatic Temporal Modulation

Another option to create a polarimeter over a large spectral range is to modulate the polarization temporally instead of spatially. However, it is impossible to build achromatic plates over such a spectral range. Therefore, it is better to define the extraction efficiency of the Stokes parameters and to achromatize these efficiencies, instead of the retardance [9] [10].

We can define a total Mueller matrix $M_{tot}$ of the polarimeter that links the input and output Stokes vector $S = (I \quad Q \quad U \quad V)^T$ by

$$S_{out} = M_{tot} \cdot S_{in} \qquad (8)$$

This total instrumental Mueller matrix is in our case only composed of the modulator's Mueller matrix and the polarizer's Mueller matrix, as the polarimeter is placed at the polarization-free Cassegrain focus of the telescope.

However, as the detector is only sensitive to the intensity, the first row of the Mueller matrix $M_{tot}$ is the only interesting part. Considering a temporal modulation of n different polarization states (by rotating the modulator at n different angular positions), the n × 4 modulation matrix $O$ is built with the first rows of the total Mueller matrices $M_{tot}$ of the n-states. It computes the intensity column vector $I'$ by

$$I' = O \cdot S_{in} \qquad (9)$$

For the polarimeter of UVMag we can use a stack of 4 birefringent $MgF_2$ plates that can rotate at 6 different angular positions [0°, 30°, 60°, 90°, 120°, 150°]. In this way, the modulation matrix $O$ is a 6 × 4 matrix and the Stokes vector measurement is over-determined. The last step of this matrices' algebra is to define the demodulation matrix $D$; [11] showed that the most optimal one is the Moore-Penrose pseudo-inverse of $O$:

$$S_{in} = D \cdot I'  \quad (10)$$

$$\text{with } D = (O^T \cdot O)^{-1} \cdot O^T \quad (11)$$

**D** is thus a 4 × 6 matrix, where 4 is the size of the Stokes vector so the first row corresponds to the intensity, the 2$^{nd}$ row to the Q component, the 3$^{rd}$ to U and the last row to the V component of the polarization. This means that each column gives the Stokes vector for each angular position of the modulator. We define extraction efficiencies of the Stokes parameters using the elements of the matrix **D**, $\varepsilon_i$ with i=[1,2,3,4] for [I, Q, U, V]:

$$\varepsilon_i = \left( 6 \cdot \sum_{j=1}^{6} D_{ij}^2 \right)^{-1/2} \quad (12)$$

These efficiencies are defined so that the propagation of the measurement noise of the i-th Stokes parameter scales with $1/\varepsilon_i$.

The goal is now to optimize these efficiencies along the wavelength range, playing with the thicknesses and fast axis orientations of the 4 retardance plates, so that we will be able to extract the Full-Stokes vector for every wavelength of interest. The two conditions on the efficiencies $\varepsilon_I^2 \leq 1$ and $\varepsilon_Q^2 + \varepsilon_U^2 + \varepsilon_V^2 \leq 1$ tell implicitly that the optimal modulation efficiency is 1 for Stokes I and $1/\sqrt{3} \approx 57.7\%$ for Stokes Q, U and V. The 8 variables (4 thicknesses or retardance value and 4 fast axis orientations) are then numerically optimized over the whole spectral range to an ideal extraction efficiency of $1/\sqrt{3}$. This is shown on the Figure 3.

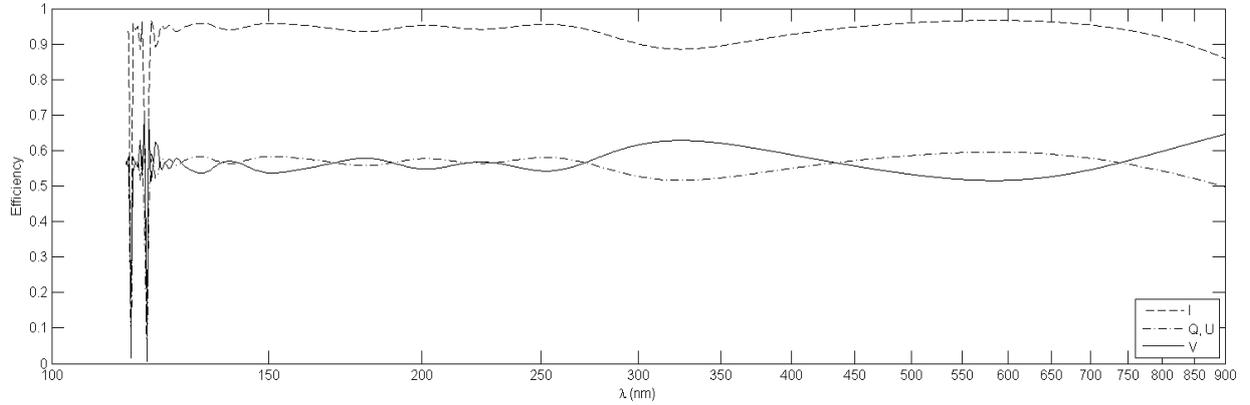

Figure 3. The optimal extraction efficiencies for the Stokes parameters are close to 57.7% for Q, U and V and close to 100% for the intensity I. The fluctuations in the FUV are due to the nulling of the birefringence.

The stack of 4 retardance plates is then followed by three Wollaston prisms in order to analyze the polarized light and to spatially separate the 2 orthogonal states. Figure 4 describes the whole polarimeter. The Wollaston prisms create a virtual slit (optically placed before the entrance of the modulator, on the left part of Figure 4), which corresponds to the entrance slit of the echelle-spectrometer, not detailed in this paper.

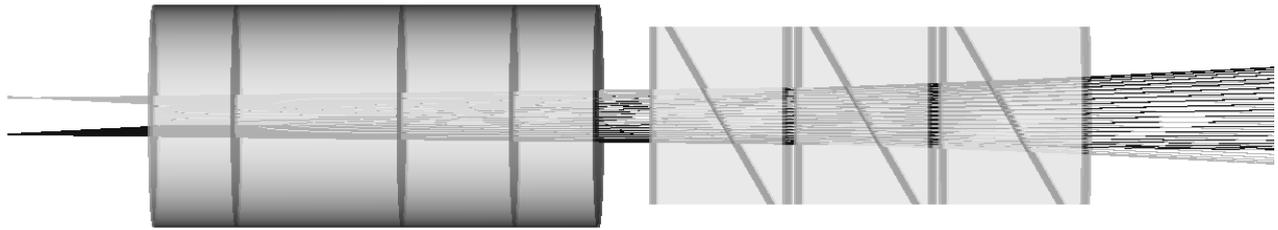

Figure 4. View of the polarimeter concept. The stack of 4 MgF$_2$ retardance plates is the modulator that takes 6 different angular positions, it has a total length of 4 mm. It is followed by the analyzer, three Wollaston prisms. The two different beams correspond to the two orthogonal polarization states separated by the Wollastons.

The angular separation between the two orthogonal polarizations is strongly dependent on the birefringence of the material used, here MgF$_2$. As shown above, the birefringence of MgF$_2$ drops dramatically below 145 nm. This means that the space gap between the two channels will be much lower in the FUV than in the visible. Another consequence is the appearance of a strong axial chromatism. The (virtual) space position of the slit will change with wavelength. To compensate this, the outside face of the last Wollaston prism is being curved in order to create a converging lens which axial chromatism will compensate the one of the modulator and Wollaston prisms.

This is a preliminary design made to show the global concept that can be used for UVMag. A more complete design will be investigated, considering the dependence on the incidence angle of the rays in the modulator, the temperature effects and taking into account the appearance of polarized fringes due to Fabry-Perot effects, [10]. These fringes are totally ignored is this very first optimization, and minimizing them may lead to extremely thin plates.

As for the first concept, some tests will be carried out in the laboratory to evaluate the possibilities offered by this method.

## 4. CONCLUSIONS AND OUTLOOK

The design of the polarimeter is the most challenging part of the UVMag instrument. The two different concepts presented in this paper are for now the most adapted for our requirements. Classical polarimeter systems are automatically eliminated because of the dispersion of their material and of the spectral range of study. The potentially best system is the second one, using a polychromatic modulator followed by Wollaston prisms. It offers a great extraction analysis of the Stokes vector over the whole spectral range with a small amount of glass for transmission reasons in the UV. The drawback of this system is that the modulator has to rotate regularly; this can be problematic as the instrument will be launched into space and require a large number of rotations during the mission lifetime. The first concept, developed by [8], has the advantage of having no moving element. But it implies the use of a relatively long entrance slit that may lead to problematically large detectors. In any case, the polarimetric performance that can be obtained with such a system is still unknown and further tests have to be carried out.

The next step for UVMag is the answer to the M4 mission call of ESA at the end of 2014. The complete optical system from the telescope to the detectors, passing through the polarimeter and the spectrograph will be developed and the calibration issues of the instrument will be determined. Meanwhile, an experimental work is planned for the two concepts of polarimeter presented in this article, in order to prove the feasibility and to determine the performances we can obtain.